\documentclass[twocolumn,showpacs,showkeys,amsmath,amssymb]{revtex4}

\usepackage{epsfig}

\begin{document}

\title{D-wave overlapping band model for cuprate superconductors}

\author{Susana Orozco}
\author{Rosa Mar\'{\i}a M\'{e}ndez-Moreno}
\author{Mar\'{\i}a de los Angeles Ortiz}
\author{Gabriela Murgu\'{\i}a}
 \email{murguia@ciencias.unam.mx}

 \affiliation{Departamento de F\'{\i}sica, Facultad de Ciencias,\\
         Universidad Nacional Aut\'{o}noma de M\'{e}xico,\\
         Apartado Postal 21-092, 04021 M\'{e}xico, D.~F., M\'{e}xico
        }

\begin{abstract}

 Within the BCS framework a multiband model with d-wave symmetry
 is considered. Generalized Fermi surface topologies via band
 overlapping are introduced. The band overlap scale is of the
 order of the Debye energy. The order parameters and the
 pairing have d-wave symmetry. Experimental values reported for
 the critical temperatures $T_c(x)$ and the order parameters,
 $\Delta_0(x)$,  in terms of dopping $x$ are used. Numerical
 results for the coupling and the band overlapping parameters in
 terms of the doping are obtained for the cuprate superconductor
 $La_{2-x}Sr_xCuO_4$.

\end{abstract}

\pacs{74.20.Fg, 74.62.-c, 74.20.-z, 74.72.Dn}
\keywords{High $T_c$, Cuprate, d-wave, Band overlap}

\maketitle


\section{Introduction}

Measurements of angle-resolved photoemission spectroscopy
(ARPES)\cite{Zhou:05} and tunneling\cite{Lee:06}, provide enough
evidence for the relevant role of phonons in high-$T_c$
superconductivity (HTSC). Experimental data accumulated so far for the
high-$T_c$ copper-oxide superconductors have given some useful clues
to unravel the fundamental ingredients responsible for the high
transition temperature $T_c$.  However, the underlying physical
process remains unknown. In this context, it seems crucial to study
new ideas that use simplified schematic models to isolate the
mechanism(s) that generate HTSC.

Pairing symmetry is an important element toward understanding the
mechanism of high-$T_c$ superconductivity. Although early experiments
were consistent with s-wave pairing symmetry, recent experiments
suggest an anisotropic pairing behavior\cite{Deutscher:05}. For many
cuprate superconductors it is generally accepted that the pairing
symmetry is d-wave for hole-doped cuprate
superconductors\cite{Tsuei:00} as for electron doped
cuprates\cite{Liu:07}. On the other hand, recent experiments with
Raman scattering and ARPES\cite{Blumberg:02,Qazilbash:05} have shown
that the gap structure on high-$T_c$ cuprate superconductors, as a
function of the angle, is similar to a d-wave
gap\cite{Hawthorn:07,Tacon:05}.  The small but non-vanishing isotope
effects in high-$T_c$ cuprates have been shown compatible with d-wave
superconductivity\cite{franck:94}. A phonon-mediated d-wave BCS like
model has recently been presented to describe layered cuprated
superconductors\cite{Xiao:07}. The last model account well for the
magnitudes of $T_c$ and the oxygen isotope exponent of the
superconductor cuprates. Calculations with BCS theory and van Hove
scenario have also been done with d-wave pairing\cite{Hassan:02}.  The
validity of d-wave BCS formalism in high-$T_c$ superconductor cuprates
has been supported by measurements of transport properties and
ARPES\cite{Matsui:05}.

Numerous indications point to the multiband nature of the
superconductivity in doped cuprates. The agreement of the multiband
model with experimental findings, suggests that a multiband pairing is
an essential aspect of cuprate superconductivity\cite{Kristoffel:08}.

First principle calculations show overlapping energy bands at the
Fermi level\cite{to}. The short coherence length observed in
high-$T_c$ superconductors, has been related to the presence of
overlapping energy bands\cite{okoye:99, saleb:08}. A simple
model with generalized Fermi surface topologies via band overlapping
has been proposed based on indirect experimental evidence. That
confirms the idea that the tendency toward superconductivity can be
enhanced when the Fermi level lies at or close to the energy of a
singularity in the density of states (DOS)\cite{Moreno:96}. This model
that can be taken as a minimal singularity in the density of states
and the BCS framework, can lead to higher $T_c$ values than those
expected from the traditional phonon barrier. In our model, the energy
band overlapping, modifies the DOS near the Fermi level allowing the
high $T_c$ values observed. A similar effect can be obtained with
other mechanisms as a van Hove singularity in the density of
states\cite{misho:2005}.

The high-$T_c$ copper-oxide superconductors have a characteristic
layered structure: the $Cu O_2$ planes. The charge carriers in these
materials are confined to the two dimensional (2D) $Cu O_2$
layers\cite{harshman:92}. These layering structures of high-$T_c$
cuprates suggest that two-dimensional physics is important for these
materials\cite{Xiao:07}.

In this work, within the BCS framework, a phonon mediated d-wave model
is proposed. The gap equation (with d-wave symmetry) and
two-dimensional generalized Fermi surface topologies via band
overlapping are used as a model for HTSC. A two overlapping band model
is considered as a prototype of multiband superconductors. For
physical consistency, an important requirement of the model is that
the band overlapping parameter is not larger than the cutoff Debye
energy, $E_D$. The model here proposed will be used to describe some
properties of the cuprate superconductor $La_{2-x}Sr_xCuO_4$ in terms
of the doping and the parameters of the model.

\section{The model}

We begin with the famous gap equation
\begin{equation}
\label{eq:az}
  \Delta(k{^\prime})= {\sum_k} V(k,k{^\prime}) 
                     \Delta(k)\frac{\tanh( E_k/2 k_B T )}{2 E_k} ,
\end{equation}
in the weak coupling limit, with $V(k,k{^\prime})$ the pairing
interaction, $k_B$ is the Boltzman constant, and $E^2_k = \epsilon^2_k
+ \Delta^2_k$, where $\epsilon_k = \hbar^2 k^2/ 2 m$ are the
self-consistent single-particle energies.

For the electron-phonon interaction, we have considered, with $V_0$ a
constant, $V(k,k^{\prime}) = V_0 \psi(k) \psi(k^{\prime})$ when
$|\epsilon_k|$ and $|\epsilon_{k^{\prime}}|~ \leq E_D ~=~k_B
T_D$ and $0$ elsewhere. As usual the attractive BCS interaction is
nonzero only for unoccupied orbitals in the neighborhood of the Fermi
level $E_F$. In the last equation, $\psi(k) = \cos ({2 ~\phi_k})$ for
$d_{x^2 -y^2}$ pairing. Here $\phi_k = \tan^{-1}(k_y / k_x)$ is the
angular direction of the momentum in the $ab$ plane. The
superconducting order parameter, $\Delta(k) = \Delta(T)~\psi(k)$ if
$|\epsilon_k| \leq E_D$ and $0$ elsewhere.

With these considerations we propose a generalized Fermi surface.  The
generalized Fermi sea proposed consists of two overlapping bands. As a
particular distribution with anomalous occupancy in momentum space the
following form for the generalized Fermi sea has been considered
\begin{equation}
\label{eq:aa}
   n_k =   \Theta(\gamma k_F - k) 
         + \Theta(\gamma k_F - k) \Theta(k - \beta k_F ),
\end{equation}
with $k_F$ the Fermi momentum and $0 < \beta < \gamma < 1$.  In order
to keep the average number of electron states constant, the parameters
are related in the 2D system by the equation
\begin{equation}
\label{eq:gg}
    2 \gamma^2 - \beta^2 = 1,
\end{equation}
then only one of the relevant parameters is independent. The
distribution in momentum induces one in energy, $E_{\beta} <
E_{\gamma}$ where $E_{\beta} = \beta^2 E_F$ and $E_{\gamma} = \gamma^2
E_F$ . We require that the band overlapping be of the order or smaller
than the cutoff (Debye) energy, which means $(1 - \gamma^2) E_F \leq
E_D$. The last expression can be written as
\begin{equation}
\label{eq:rr}
    (1 - \gamma^2) E_F = \eta  E_D,
\end{equation}
where $\eta$ is in the range $0 < \eta < E_F/( 2 E_D)$.  Equations
(\ref{eq:gg}) and (\ref{eq:rr}) together will give the minimum
$\gamma^2$ value consistent with our model.

In the last framework the summation in Eq.~(\ref{eq:az}) is changed to
an integration which is done over the ({\it symmetric}) generalized
Fermi surface defined above. One gets
\begin{equation}
\label{eq:bb}
\begin{split}
 1  = & ~\frac{\lambda}{4\pi} \int_{E_\gamma - E_D}^{E_\gamma + E_D}
           \int_{0}^{ 2\pi} d\phi~\cos^2({2\phi}) 
               \tanh \left(\frac{\sqrt{\Xi_k}}{2 k_B T}\right)
               \frac{d\epsilon_k}{\sqrt{\Xi_k}} \\
      & + \frac{\lambda}{4\pi} \int_{E_\beta}^{ E_F} 
             \int_{0}^{ 2\pi} d{\phi}~\cos^2({2\phi}) 
                 \tanh \left(\frac{\sqrt{\Xi_k}}{2 k_B T }\right)
                 \frac{d\epsilon_k}{\sqrt{\Xi_k}}.
\end{split}
\end{equation}

In this equation $\Xi_k = (\epsilon_k - E_F)^2 + \Delta(T)^2
~\cos^2({2~\phi})$, the coupling parameter is $\lambda = V_0 D(E)$,
with $D(E)$ the electronic density of states, which will be taken as a
constant for the $2D$ system in the integration range. $E_F~=
\frac{{\hbar}^2\pi}{m}n_{2D}$, with $n_{2D}$ the carriers density per
$CuO_2$ layer. The two integrals correspond to the bands proposed by
Eq.~(\ref{eq:aa}).

The integration over the surface at $E_{\gamma}$ in the first band, is
restricted to states in the interval $E_{\gamma} - E_D \leq E_k \leq
{E_{\gamma}+ E_D}$. In the second band, in order to conserve the
particle number, the integration is restricted to the interval
$E_{\beta} \leq E_k \leq {E_F}$, if ~$E_{\gamma}+ E_D>E_F$, with
$E_{\beta} ~=~ (2~\gamma^2 ~-~1 ) E_F$, according to Eq.~(\ref{eq:gg})
in our model. While ~ $E_F - E_{\gamma} \leq E_D$, implies that the
energy difference between the anomalously occupied states must be
provided by the material itself. Finally $ \Delta(T)~\psi(k) =
\Delta(T)~\cos({2~\phi})$ at the two bands.

The critical temperature is introduced via the Eq.~(\ref{eq:bb}) at $T
= T_c$, where the gap becomes $\Delta(T_c) = 0$. At this temperature
Eq.~(\ref{eq:bb}) is reduced to
\begin{equation}
\begin{split}
\label{eq:cc}
1 = & ~\frac{\lambda}{4}~ \int_{E_\gamma - E_D}^{E_\gamma + E_D}
               \tanh \left(\frac{\epsilon_k - E_F}{2 k_B T_c}\right)
               \frac{d\epsilon_k}{\epsilon_k - E_F}  \\
    & +       \frac{\lambda}{4}~ \int_{E_\beta}^{E_F} 
               \tanh \left(\frac{\epsilon_k - E_F}{2 k_B T_c}\right)
               \frac{d\epsilon_k}{\epsilon_k - E_F},
\end{split}
\end{equation}
which will be numerically evaluated. The last equation relates $T_c$
to the coupling constant $\lambda$ and to the anomalous occupancy
parameter $\gamma^2$. This relationship determines the $\gamma^2$
values which reproduces the critical temperature of several cuprates
in the weak coupling region.

At $T = 0$K, Eq.~(\ref{eq:bb}) will also be evaluated and $\gamma^2$
values consistent with the numerical results of Eq.~(\ref{eq:cc})
will be obtained:
\begin{equation}
\label{eq:ee}
\begin{split}
 1~ = & ~\frac{\lambda}{4~\pi}~\int_{0}^{ 2\pi} d\phi~\cos^2({2~\phi}) \\
     & \times ~\left[
            \sinh^{-1}~ \frac{~k_B T_D~-~(1~-~\gamma^2)k_B T_F}
                            {\Delta_0~|\cos{(2~\phi)}|}~ \right. \\
         & \qquad + ~\sinh^{-1}~ \frac{~(1~-~\gamma^2)k_B T_F ~+ ~k_B T_D}
                                   {\Delta_0~|\cos({2~\phi})|}\\
         & \qquad + ~\left. \sinh^{-1}~\frac{2k_B~(1~-~\gamma^2)T_F}
                             {\Delta_0~|\cos{(2~\phi)}|}
          \right],
\end{split}
\end{equation}
where $\Delta(0) = \Delta_0$.

The model presented in this section can be used to describe high-$T_c$
cuprate superconductors, the band overlapping $1~-~\gamma^2$ and
relevant parameters are determined. In any case a specific material
must be selected to introduce the available experimental data. Ranges
for the coupling parameter $\lambda$ in the weak coupling region, and
the overlapping parameter $\gamma^2$, consistent with the model and
the experimental data, can be obtained for each material. The
relationship between the characteristic parameters will be obtained
for $La$-based compounds at several doping concentrations $x$, ranging
from the underdoped to the overdoped regime. Different values of the
coupling constant and the overlapping parameter consistent with the
model, are obtained using the experimental values of $\Delta_0$ and
$T_c$.

The single layer cuprate superconductor $La_{2-x}Sr_xCuO_4$ ($La-214$)
has one of the simplest crystal structures among the high-$T_c$
superconductors. This fact makes this cuprate very attractive for both
theoretical and experimental studies. High quality single crystals of
this material are available with several doping concentrations which
are required for experimental studies.  Even the determination of
charge carrier concentration in the cuprate superconductors is quite
difficult, the $La-214$ is a system where the carrier concentration is
nearly unambiguously determined.  For this material, the hole
concentration for $CuO_2$ plane, $n_{2D}$, is equal to the $x$ value,
{\it i.e.} to the $Sr$ concentration, as long as the oxygen is
stoichiometric\cite{Ando:00,Ino:02}. Additionally, there are
reliable data for the $T_c$ and the superconducting gap $\Delta_0$ for
several samples in the superconducting region.

\section{Results and discussion}

In order to get numerical results, with our overlapping band model
with d-wave symmetry, the cuprate $La_{2-x} Sr_x Cu O_4$ was
selected. The values for $\Delta_0$ are taken in the interval $2 \leq
~ \Delta_0 ~\leq 12$~meV which includes experimental
results\cite{Ino:02}. The behavior of $\lambda$ as function of $x$ and
$\gamma^2$ at $T = T_c$ is obtained from Eq.~(\ref{eq:cc}); and
$\lambda$ as function of $\Delta_0$, $x$ and $\gamma^2$ at $T = 0$K is
given by Eq.~(\ref{eq:ee}). To have coupled solutions of these
equations the same $\lambda$ value for $T = T_c$ and $T = 0$K is
proposed. These solutions correspond to different overlap values $1 -
\gamma^2$, at each equation. With this model and s-wave symmetry, the
band overlapping $1 - \gamma^2$ was higher at $T = 0$K than at $T =
T_c$\cite{oro:07}. We consider the same behavior with d-wave
symmetry. The maximum $T_c$ for cuprate superconductors is obtained at
optimal doping. With the model $\lambda(x)$ values are obtained,
including at optimal doping $\lambda(x_{op})$\cite{oro:08}.

\begin{center}
\begin{figure}[h]
\epsfig{file=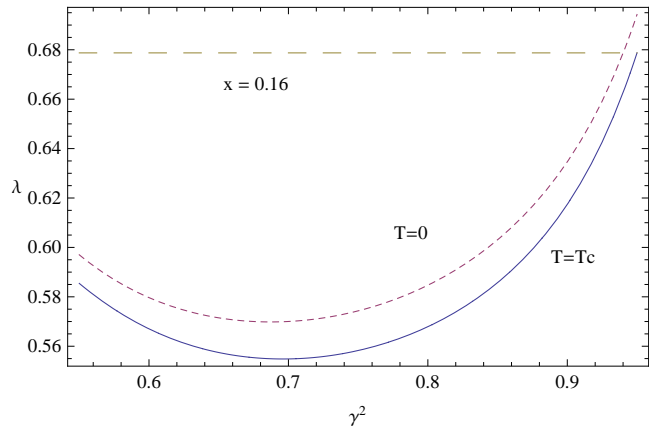,bb=0 0 240 160,angle=0}
 \caption[The coupling parameter $\lambda$ in terms of the overlapping
   parameter $\gamma^2$, with optimal doping $x= 0.16$. The $T = 0$K
   (broken curve) and $T = T_c = 40$K (continuous curve) for
   $La_{2-x} Sr_x Cu O_4$, are shown. The horizontal line at $\lambda
   = 0.68$ shows the maximum $\lambda$ selected for $T = T_c$ with
   $\gamma^2 = 0.95$.]
   {The coupling parameter $\lambda$ in terms of the overlapping
     parameter $\gamma^2$, with optimal doping $x= 0.16$. The $T = 0$K
     (broken curve) and $T = T_c = 40$K (continuous curve) for
     $La_{2-x} Sr_x Cu O_4$, are shown. The horizontal line at
     $\lambda = 0.68$ shows the maximum $\lambda$ selected for $T =
     T_c$ with $\gamma^2 = 0.95$.}
 \label{fig1}
\end{figure}
\end{center}

In Fig.~\ref{fig1}. values of the coupling parameter $\lambda$ in terms of the
overlapping parameter $\gamma^2$ are shown in the weak coupling
region. The experimental results of $T_c$ and $\Delta_0$ from
Refs.~\cite{harshman:92} and~\cite{Ino:02} were introduced. The curves
at $T = 0$K (broken curve) and at $T = T_c = 40$K (continuous curve)
for $La_{2-x} Sr_x Cu O_4$, with optimal doping $x_{op} = 0.16$ are
shown. The minimum $\gamma^2$ value of $0.55$ was taken to be
consistent with the model. In the whole range reported for the band
overlapping, the coupling parameter required at each $\gamma^2$ is
larger for $T = 0$K than for $T = T_c$. In order to use the same
$\lambda$ for $T = T_c$ and $T = 0$K, the $\lambda$ values must be
restricted {\it i.e.}, the $\lambda$ value at each $\gamma^2$ must be
larger than $\lambda_{min}= 0.57$ at $T = 0$K.

In the region $\gamma^2 \geq 0.7$ with a constant $\lambda$ value, a
larger band overlapping $1 - \gamma^2$ is obtained for $T = 0$K than
for $ T = T_c$ in agreement with our assumption.  For instance, the
maximum $\lambda$ for $T = T_c$ with $\gamma^2 = 0.95$, is shown by
the horizontal line at $\lambda = 0.68$, and the intersection of this
line and the $T = 0$K curve is at $\gamma^2 = 0.94$.  The same
restrictions over $\lambda$ are considered at any other doping in the
superconducting phase. However, for any $x \neq x_{op}$, the $\lambda
$ value must be smaller than $\lambda = 0.68$.

\begin{center}
\begin{figure}[h]
\epsfig{file=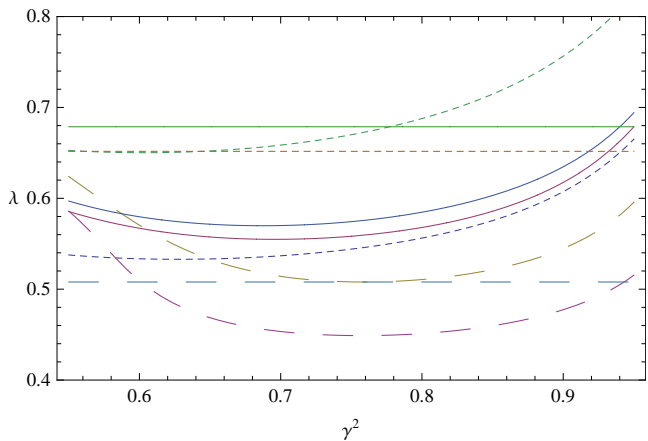,bb=0 0 240 166,angle=0}
 \caption[The results for optimal doping (continuous curves) are
   compared with the underdoped (small dashed curves) $ x= 0.13$ and
   the overdoped $x= 0.2$ (large dashed curves) cases. The three
   horizontal lines show the selected extreme $\lambda$ values:
   $\lambda = 0.68$ at optimal doping, $\lambda= 0.65$ in the
   underdoped case, and $\lambda = 0.51$ in the overdoped case.]
   {The results for optimal doping (continuous curves) are compared
     with the underdoped (small dashed curves) $ x= 0.13$ and the
     overdoped $x= 0.2$ (large dashed curves) cases. The three
     horizontal lines show the selected extreme $\lambda$ values:
     $\lambda = 0.68$ at optimal doping, $\lambda= 0.65$ in the
     underdoped case, and $\lambda = 0.51$ in the overdoped case.}
 \label{fig2}
\end{figure}
\end{center}

In Fig.~\ref{fig2} the results for optimal doping $ x_{op}$, are compared with
the underdoped $x=0.13$ and the overdoped $x=0.2$ cases. The
experimental values of $\Delta_0$ and $T_c$ for each doping, were
introduced. The continuous curves correspond to $ x_{op}$, the small
dashed curves show the underdoped behavior and the large dashed ones
the overdoped results. In the optimal doped and underdoped cases, the
$T = 0$K curves are above the corresponding $T = T_c$ ones. In the
overdoped case, the behavior is different {\it i.e.}, the $T_c$ curve
is above the $T = 0$K one.

In the three cases, the values of the coupling parameter are in the
weak coupling region for the $\gamma^2$ values which satisfy the
conditions of our model. All the $\gamma^2$ values which satisfy the
$\lambda$ restrictions are allowed. However, as an example, we have
selected extreme $\lambda$ values in the three cases. The three
horizontal lines show these $\lambda$ values.

As in Fig.~\ref{fig1} the maximum $\lambda $ value selected at optimal
doping is $\lambda = 0.68$. In the underdoped case $\lambda= 0.65$ is
selected. This value corresponds to the overlapping parameter
$\gamma^2= 0.621$, the minimum of the $T = 0$K curve, and $\gamma^2=
0.941$ at the $T = T_c$ curve. In the overdoped case, the selected
$\lambda$ value is $ 0.51$, {\it i.e.} the minimum of the curve $T =
T_c$. With this $\lambda$, the overlapping parameters are $\gamma^2=
0.599$ for $T = 0$K and $\gamma^2= 0.76$ for $T=T_c$.

\begin{center}
\begin{figure}[ht]
\epsfig{file=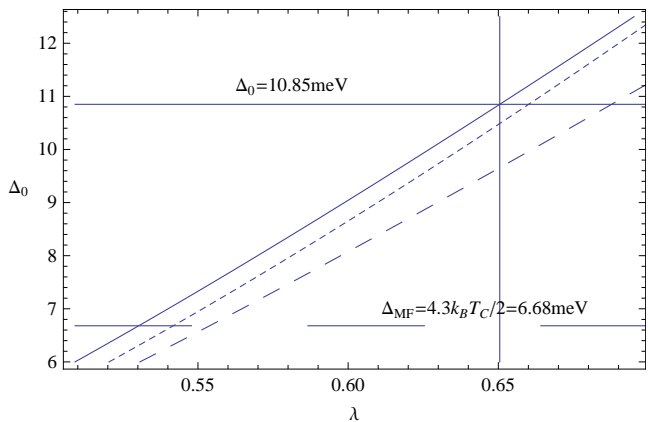,bb=0 0 240 157,angle=0}
 \caption[The gap $\Delta_0$ obtained in terms of the coupling
   parameter $\lambda$ for the underdoped sample. The curves are drawn
   for $\gamma^2 = 0.621, 0.5$ and $0.8$ from up to down
   respectively. The continuous horizontal line shows the experimental
   $\Delta_0 = 10.85$meV value. The large dashed horizontal line
   shows the d-wave mean-field approximation $\Delta_{MF}= 6.68$meV
   result.]
   {The gap $\Delta_0$ obtained in terms of the coupling parameter
     $\lambda$ for the underdoped sample. The curves are drawn for
     $\gamma^2 = 0.621, 0.5$ and $0.8$ from up to down
     respectively. The continuous horizontal line shows the
     experimental $\Delta_0 = 10.85$meV value. The large dashed
     horizontal line shows the d-wave mean-field approximation
     $\Delta_{MF}= 6.68$meV result.}
 \label{fig3}
\end{figure}
\end{center}

With numerical solutions of Eq.~(\ref{eq:ee}) we may obtain the gap
$\Delta_0$ in terms of the parameters of our model. The underdoped
material is considered in Fig.~\ref{fig3} because the advantage of our
model is easily shown. The gap $\Delta_0$ is shown in terms of the
coupling parameter $\lambda$. The gap $\Delta_0$ always increases with
the coupling parameter $\lambda$. The curves are drawn for $\gamma^2 =
0.621, 0.5$ and $0.8$ from up to down respectively. For this sample,
with $\gamma^2 = 0.621$, we obtain the minimum $\lambda$ value for any
$\Delta_0$ and for any $\lambda$ the maximum $\Delta_0$ value.

The continuous horizontal line shows the experimental $\Delta_0 =
10.85$meV value. The large dashed horizontal line shows the d-wave
mean-field approximation $\Delta_{MF}= 6.68$meV result\cite{won:94},
where the same d-wave symmetry was considered.  However, introduction
of the band overlapping allows to reproduce the experimental result
with all the $\gamma^2$ values in the range considered. The band
overlapping model also allows higher $\Delta_0$ values for the
underdoped system and lower $\Delta_0$ for the overdoped one, than the
$\Delta_{MF}= 2.145 k_B T_c$.

\begin{center}
\begin{figure}[ht]
\epsfig{file=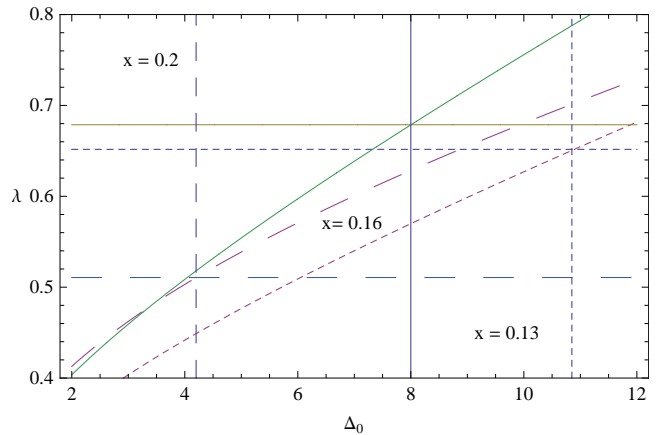,bb=0 0 240 164,angle=0}
 \caption[$\Delta_0$ as function of $\lambda$ for optimal doping,
   underdoped, and the overdoped cases. The horizontal lines are the
   $\lambda$ values selected in Fig.~\ref{fig2}. The vertical lines show the
   experimental $\Delta_0$ values. The continuous lines correspond to
   $x = 0.16$, the small dashed ones to $x = 0.13$, and the large
   dashed ones to $x = 0.2$.]
   {$\Delta_0$ as function of $\lambda$ for optimal doping,
     underdoped, and the overdoped cases. The horizontal lines are the
     $\lambda$ values selected in Fig.~\ref{fig2}. The vertical lines show the
     experimental $\Delta_0$ values. The continuous lines correspond
     to $x = 0.16$, the small dashed ones to $x = 0.13$, and the large
     dashed ones to $x = 0.2$.}
 \label{fig4}
\end{figure}
\end{center}

In Fig.~\ref{fig4} the behavior between $\Delta_0$ and $\lambda$ for
optimal doping is compared with the underdoped and the overdoped
cases. The $\gamma^2$ values introduced are those selected in
Fig.~\ref{fig2} for $T = 0$K. The horizontal lines are the $\lambda$
values also selected in Fig.~\ref{fig2}. All the continuous curves
correspond to optimal doping. The large and small dashed curves
correspond to the overdoped and the underdoped systems
respectively. The curves show the interesting relationship between
these parameters. As for optimal doping, the coupling parameter
increases with $\Delta_0$ for any doping. The vertical lines are the
experimental $\Delta_0$ values. It is possible to reproduce the
experimental $\Delta_0$ in the range $0.13 \leq x \leq 0.2$. The band
overlapping introduced in this model allows the reproduction of the
behavior of $\Delta_0$ with doping.

In conclusion, we presented an overlapping band model with d-wave
symmetry, to describe high-$T_c$ cuprate superconductors, within the
BCS framework. We have used a model with anomalous Fermi Occupancy and
d-wave pairing in the 2D fermion gas. The anomaly is introduced via a
generalized Fermi surface with two bands as a prototype of bands
overlapping.  We report the behavior of the coupling parameter
$\lambda$ as function of the gap $\Delta_0$ and the overlapping
parameter $\gamma^2 $, for different doping samples. The $\lambda$
values consistent with the model are in the weak coupling region.  The
behavior of $\Delta_0$ as function of $\lambda$ shows that for several
band overlapping parameters it is possible to reproduce the
experimental $\Delta_0$ values near the optimal doping, for the
cuprate $La_{2-x} Sr_x Cu O_4$. The band overlapping allows the
improvement of the results obtained with a d-wave mean-field
approximation, in a scheme in which the electron-phonon interaction is
the relevant high-$T_c$ mechanism. The energy scale of the
anomaly $(1 - \gamma^2)E_F$ is of the order of the Debye energy. The
Debye energy is then the overall scale that determines the highest
$T_c$ and gives credibility to the model because it requires an energy
scale accessible to the lattice. The enhancing of the DOS with this
model simulates quite well intermediate and strong coupling
corrections to the BCS framework.



\end{document}